\documentclass[aps,preprint,superscriptaddress,prl,floatfix]{revtex4}

\usepackage{graphicx}
\usepackage{amsmath}
\usepackage{amssymb}
\usepackage{units}
\usepackage{times}

\begin{document}

%
%

\title{Spin and angular resolved photoemission experiments on epitaxial graphene}
\author{Isabella Gierz}
\email[Corresponding author; electronic address:\
]{i.gierz@fkf.mpg.de} \affiliation{Max-Planck-Institut f\"ur
Festk\"orperforschung, D-70569 Stuttgart, Germany}
\author{Jan Hugo Dil}
\affiliation{Swiss Light Source, Paul Scherrer Institut, CH-5232
Villigen, Switzerland}\affiliation{Physik-Institut, Universit\"at
Z\"urich, CH-8057 Z\"urich, Switzerland}
\author{Fabian Meier}
\affiliation{Swiss Light Source, Paul Scherrer Institut, CH-5232
Villigen, Switzerland}\affiliation{Physik-Institut, Universit\"at
Z\"urich, CH-8057 Z\"urich, Switzerland}
\author{Bartosz Slomski}
\affiliation{Swiss Light Source, Paul Scherrer Institut, CH-5232
Villigen, Switzerland}\affiliation{Physik-Institut, Universit\"at
Z\"urich, CH-8057 Z\"urich, Switzerland}
\author{J\"urg Osterwalder}
\affiliation{Physik-Institut, Universit\"at Z\"urich, CH-8057
Z\"urich, Switzerland}
\author{J\"urgen Henk}
\affiliation{Max-Planck-Institut f\"ur Mikrostrukturphysik,
D-06120 Halle (Saale), Germany}
\author{Roland Winkler}
\affiliation{Department of Physics, Northern Illinois University, USA}
\author{Christian R. Ast}
\affiliation{Max-Planck-Institut f\"ur Festk\"orperforschung,
D-70569 Stuttgart, Germany}
\author{Klaus Kern}
\affiliation{Max-Planck-Institut f\"ur Festk\"orperforschung,
D-70569 Stuttgart, Germany}\affiliation{Institut de Physique de la
Mati{\`e}re Condens{\'e}e, Ecole Polytechnique F{\'e}d{\'e}rale de
Lausanne, CH-1015 Lausanne, Switzerland}
\date{\today}

\begin{abstract}
Our recently reported spin and angular resolved photoemission (SARPES) results on an epitaxial graphene monolayer on SiC(0001) suggested the presence of a large Rashba-type spin splitting of $\Delta k=(0.030\pm0.005)$\AA$^{-1}$ \cite{Gierz2010}. Although this value was orders of magnitude larger than predicted theoretically, it could be reconciled with the line width found in conventional spin-integrated high resolution angular resolved photoemission spectroscopy (ARPES) data. Here we present novel measurements for a hydrogen intercalated quasi free-standing graphene monolayer on SiC(0001) that reveal a spin polarization signal that --- when interpreted in terms of the Rashba-Bychkov effect \cite{Bychkov1984_1,Bychkov1984_2} --- corresponds to a spin splitting of $\Delta k=(0.024\pm0.005)$\AA$^{-1}$. This splitting is significantly larger than the half width at half maximum of spin-integrated high resolution ARPES measurements which is a strong indication that the measured polarization signal does not originate from a Rashba-type spin splitting of the graphene $\pi$-bands as we suggested in our previous report \cite{Gierz2010}.
\end{abstract}

\maketitle

\section{sample preparation}
Hydrogen intercalated graphene monolayers on SiC(0001) were prepared as described in Ref. \cite{Riedl2009}. Prior to graphitization in an induction furnace under 900 mbar of argon \cite{Emtsev2009}, the SiC(0001) crystal was hydrogen etched to remove scratches caused by mechanical polishing \cite{Owman1996}. The graphitization process was interrupted after the formation of the first carbon monolayer (the so-called buffer layer) which was subsequently decoupled from the SiC(0001) substrate by hydrogen intercalation resulting in a quasi free-standing graphene monolayer \cite{Riedl2009}.

\section{high resolution ARPES measurements}
\begin{figure}
  \includegraphics[width = 1\columnwidth]{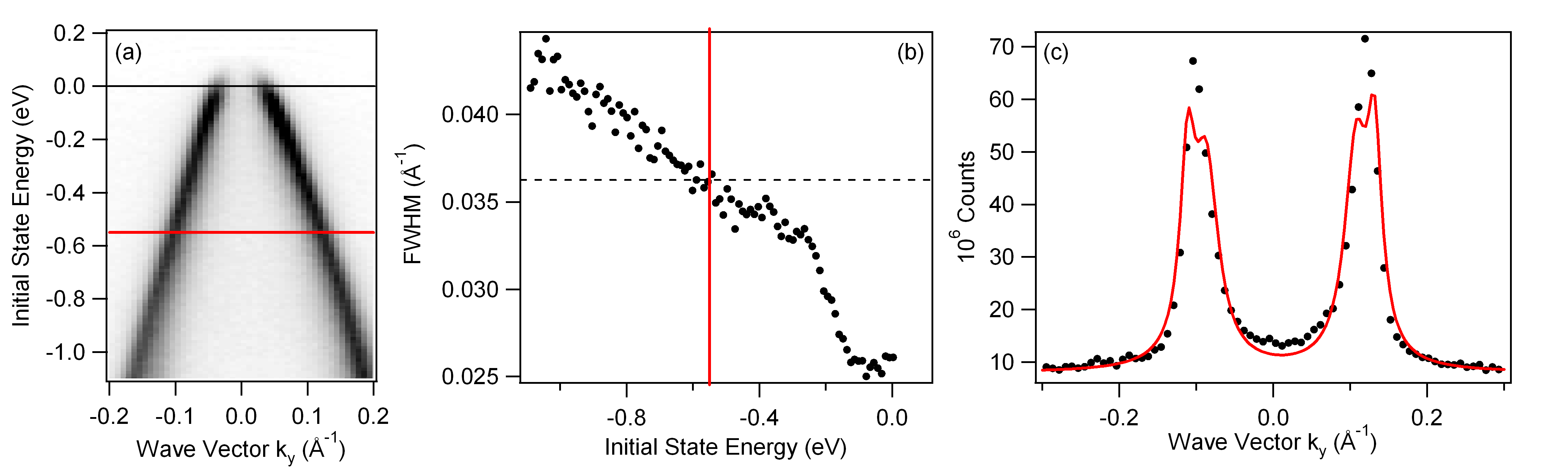}
  \caption{Panel (a) displays the $\pi$-band dispersion for a hydrogen intercalated graphene monolayer on SiC(0001). The corresponding full width at half maximum (FWHM) of the $\pi$-bands as a function of initial state energy is shown in panel (b). The attempt to fit the spin-integrated momentum distribution curve taken at an initial state energy of $-$0.55\,eV (see red line in panel a) with two components per peak separated by 0.024\AA$^{-1}$ fails (c).}
  \label{figure1}
\end{figure}

High-resolution ARPES experiments were performed with an energy and angular resolution of 10\,meV and 0.5$^{\circ}$, respectively. The sample was transferred to the ARPES chamber in air and subsequently cleaned by a mild annealing at around 200$^{\circ}$C. Figure \ref{figure1} (a) shows the band structure of the H-intercalated graphene monolayer around the $\overline{\text{K}}$-point of the two-dimensional Brillouin zone measured perpendicular to the $\overline{\Gamma\text{K}}$-direction along $k_y$. We observe two linearly dispersing $\pi$-bands that cross approximately 0.33\,eV above the Fermi level, i.e. the hydrogen intercalated graphene monolayer is slightly p-doped. We determined the full width at half maximum (FWHM) of the $\pi$-bands by fitting momentum distribution curves (MDCs) with Lorentzian line shapes and a constant background. The FWHM as a function of the initial state energy is shown in Fig. \ref{figure1} (b). The line width of the H-intercalated graphene monolayer is significantly smaller than the line width of the graphene sample prepared by graphitization of SiC(0001) in ultra high vacuum that we investigated in our previous report \cite{Gierz2010}.

\section{spin-resolved measurements}
\begin{figure}
  \includegraphics[width = 1\columnwidth]{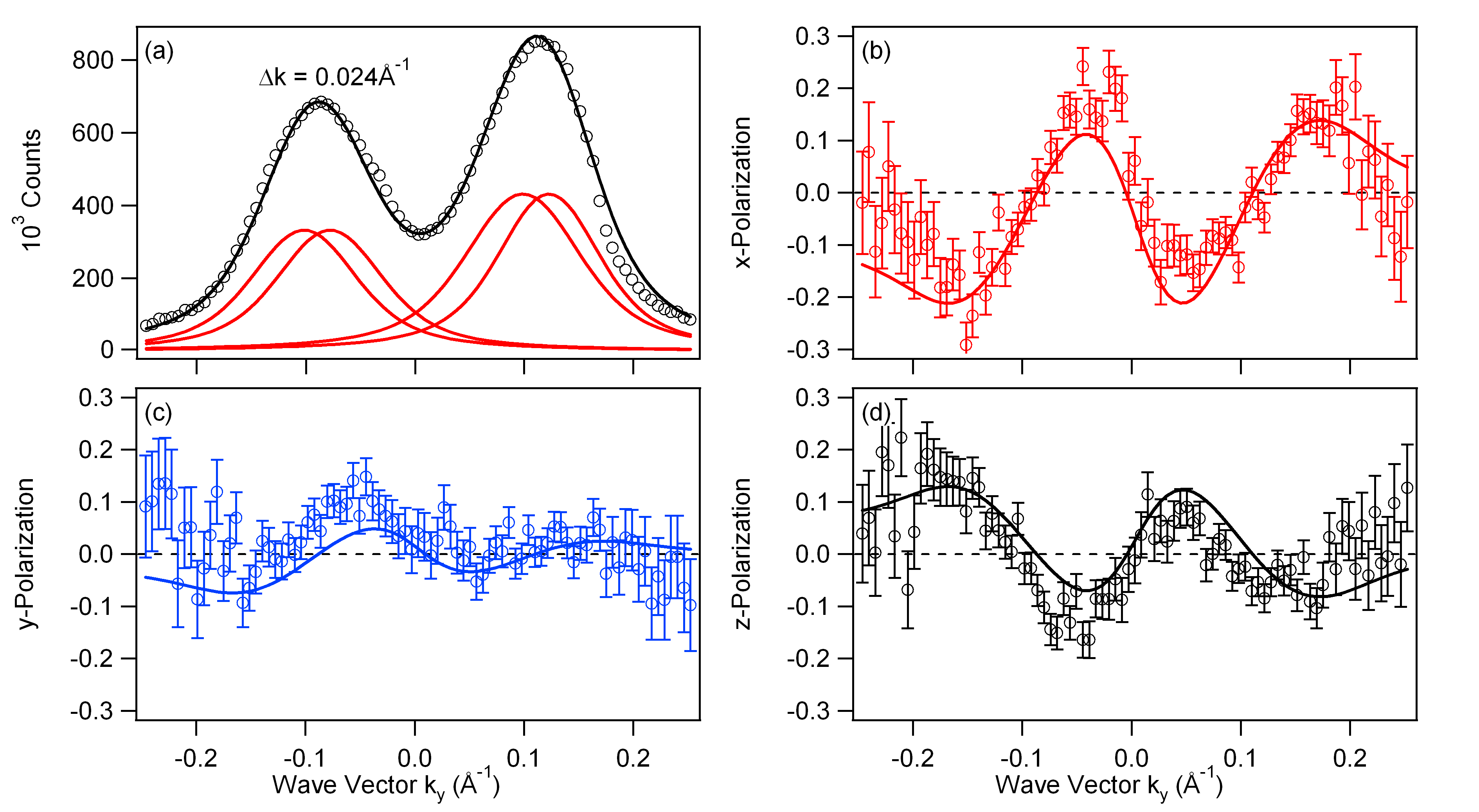}
  \caption{Panels (b)-(d) show the measured spin polarization for the corresponding spin-integrated momentum distribution curve in panel (a). The statistical error bars in this figure were calculated via $\sqrt{I_{tot}}/S$, where $I_{tot}$ is the total intensity and $S$ is the experimental Sherman function.}
  \label{figure2}
\end{figure}

Spin-resolved measurements were taken with the COPHEE setup at the Swiss Light Source with an energy and momentum resolution of 80\,meV and 4$^{\circ}$, respectively, for a photon energy of 30\,eV. Again, the sample was transferred to the SARPES chamber in air and subsequently cleaned by a mild annealing at around 200$^{\circ}$C. Similar to our previous spin-resolved ARPES experiments \cite{Gierz2010} we measured the spin-polarization of the photoelectrons along all three spatial directions for an MDC at an initial state energy of $-$0.55\,eV [see red line in Fig. \ref{figure1} (a)]. These measurements were made by rotating the sample around the surface normal (azimuthal scan), effectively crossing the $\pi$-bands in the direction perpendicular to the plane spanned by the direction of incidence of the light and the detection direction of the electrons. Figure \ref{figure2} (b)-(d) shows the $x$, $y$ (in-plane) and $z$ (out-of-plane) components of the measured spin-polarization along $k_y$ (perpendicular to the $\overline{\Gamma\text{K}}$-direction). The polarization signal looks similar to what is expected for a Rashba-type spin splitting of the initial state. Therefore, one might be tempted to interpret the measured polarization
in terms of a Rashba splitting of the $\pi$-bands. We analyzed the data set displayed in Fig. \ref{figure2} accordingly, using
the two-step fitting routine from Ref. \cite{Meier2008}. From the measured spin-polarization we may conclude that the spin-integrated MDC shown in Fig. \ref{figure2} (a) actually consists of four spin-polarized peaks that are shown as red lines. From these fits we obtain a spin splitting of $\Delta k=(0.024\pm0.005)$\AA$^{-1}$. \\

A comparison with the FWHM of the $\pi$-bands in Fig. \ref{figure1} (b) reveals that the fit result for a possible spin splitting is significantly larger than the half width at half maximum of the $\pi$-bands (0.018\AA$^{-1}$ at $-$0.55\,eV). More precisely, it is impossible to fit the high-resolution MDC displayed in Fig. \ref{figure1} (c) with two Lorentzian components per peak separated by 0.024\AA$^{-1}$. This is a clear indication that the measured spin polarization cannot be attributed to a Rashba-type spin splitting of the graphene $\pi$-bands as we suggested earlier \cite{Gierz2010}.

\section{conclusion}

At present, we are unable to say where the measured polarization signal is coming from. However, from the measurements presented here, we conclude that the origin of the measured polarization is not a Rashba-type spin splitting of the graphene $\pi$-bands. Therefore, we hereby explicitly revoke our previous interpretation of the observed polarization signal in terms of a Rashba-type spin splitting of the graphene $\pi$-bands.

\section{acknowledgement}
The authors thank the group of T.\ Seyller at the University of Erlangen-N\"urnberg for the preparation of the hydrogen intercalated graphene monolayer. C.\ R.\ A.\ acknowledges funding by the Emmy-Noether-Program of the Deutsche Forschungsgemeinschaft (DFG). We are especially grateful for comments by E.\ Rotenberg concerning the possible origin of the observed polarization signal.

\end{document}